\documentclass[a4paper,12pt]{article}
\usepackage{dsfont}
\usepackage{amsmath}
\usepackage{accents}
\usepackage{bm}
\usepackage{multirow}
\usepackage[english]{babel}
\usepackage{graphicx}

\begin{document}
\title{Probing the geometry of two-qubit state space by evolution}
\author{{\bf \sf Andrzej M. Frydryszak$^1$, Maria Gieysztor$^{2,3}$ and Andrij Kuzmak$^4$  }\\[3mm]
 $^{1)}$ Institute of Theoretical Physics,\\
University of Wroclaw, pl. M. Borna 9,\\
50 - 204 Wroclaw, Poland\\
{\small \em e-mail: andrzej.frydryszak@ift.uni.wroc.pl}
\\[2mm]
$^{2)}$ Institute of Theoretical Physics,\\
University of Wroclaw, pl. M. Borna 9,\\
50 - 204 Wroclaw, Poland\\
{\small \em e-mail: maria.gieysztor@ift.uni.wroc.pl}
\\[2mm]
$^{3)}$ Faculty of Physics, Astronomy and Informatics\\ 
Nicolaus Copernicus University\\
Grudziadzka 5, 87-100 Torun, Poland\\
\\[2mm]
$^{4)}$ Department of Theoretical Physics,\\ Ivan Franko National
University of Lviv,\\ 12 Drahomanov St., Lviv, UA-79005, Ukraine\\
{\em \small  e-mail: andrijkuzmak@gmail.com}}

\maketitle

\begin{abstract}
We derive an explicit expressions for geometric description of state manifold obtained from evolution governed by
a three parameter family of Hamiltonians covering most cases related to real interacting two-qubit systems. We discuss types of evolution in terms of  the defining parameters and obtain relevant explicit description of the pure state spaces and their Remannian geometry with the Fubini-Study metric . In particular, there is given an analysis of the modification of known geometry of quantum state manifold by the linear noncommuting perturbation of the Hamiltonian.  For families of states  resulting from the unitary evolution, we characterize a degree of entanglement using the squared concurrence as its measure.
\end{abstract}
%
\section{Introduction}

The precise geometric description of the full state space of quantum system is crucial in studying its physical properties
\cite{gqev1,gqev2,gqev3,FSM2,OCGQC,GAQCLB,QCAG,QGDM,qcomp1,qcomp2,qcomp3,qcomp5,FSM4,CSRS,GQSMGLA,torus,FMM}.
Specially, for compound systems, where characterization beyond the general property of convexity, for multilevel-systems, gets rather involved, even for bipartite systems. In general, such a quantum state space 
cannot be expected to form a smooth manifold. On the other hand, there is an option to focus on distinguished subsets
of  states of quantum system,  namely, on orbits generated by unitary evolutions defined by physically relevant Hamiltonians.
Such a focus has been proven fruitful from various perspectives like, the control theory \cite{OCTUT,SRGTOC}, the quantum brachistochrone problem
\cite{OHfST,QBS1}, a time-optimal evolution \cite{OCGQC,GAQCLB,QCAG,QGDM,SRGTOC,brach,brachass},
or the question of Zermelo navigation \cite{ZNP2,ZNP3,ZNP4}.

The geometry of the set of quantum states obtained as result of a family of unitary evolutions depending on a set of 
parameters  can be naturally studied with the use of the Fubini-Study metric \cite{gqev2,FSM2,FSM3,FSM0,FSM1,FSMqsgm,FSMqsgm2}, where the dimension of such obtained 
Riemannian manifold is equall to the number of parameters. Obviously, the details of such orbits depend on applied 
Hamiltonian  and selected initial state.  Most of interactions of a two-qubit systems can be described by the generalized 
Heisenberg type interaction Hamiltonian containing anisotropic terms, which is conventionally put into the following 
physical form suitable e.g. for studying quantum dot systems \cite{LD}
\begin{equation}
H=J\left(
\vec{S}_1\cdot \vec{S}_2+\vec{\kappa}_{DM}\cdot(\vec{S}_1\times \vec{S}_2)+ \vec{S}_1\Theta \vec{S}_2
\right),
\end{equation}       
where $\vec{S}_j$ are spins of $j=1,2$ of subsystems, $\vec{\kappa}_{DM}$ is the Dzyaloshinski-Moriya vector controlling 
anti-symmetric part and $\Theta$ is symmetric, traceless $3\times 3$ matrix. However, for further considerations, to 
describe manifolds of states, we shall use the explicit $\sigma$-matrix notation. Above interaction Hamiltonian is covered 
by the general form of a nonlocal Hamitonian for two-qubit system
\begin{equation}
H=\sum_{i,j=1}^3 h_{ij}\sigma_i\otimes \sigma_j ,
\end{equation}    
$h_{ij}$ are real nad $\sigma_j$, $j=1,2,3$ are Pauli matrices. As it is known \cite{ZVSW}, such a Hamiltonian can be
transformed into the diagonal form 
\begin{equation}
H_{int}=\sum_{j=1}^3 c_j\sigma_j\otimes\sigma_j
\end{equation}
The full Hamiltonian contains additionaly a local term $H_{loc}= H_1 \otimes \mathds{1} + \mathds{1} \otimes H_2$, with 
$H_a$, $a=1,2$ being on-qubit Hamiltonians. For simplicity, in the present work we shall fix two-qubit local Hamiltonian 
in the form of the coupling of both systems to an external magnetic field along third axis i.e. 
\begin{equation}
H_{loc}=H_0 \equiv b\, ( \sigma_3 \otimes \mathds{1}+ \mathds{1}\otimes \sigma_3).
\end{equation}
In the following we shall consider unitary evolutions generated by four-parameter family of Hamiltonians
\begin{equation}\label{ham}
H(b, c_1, c_2, c_3)= H_0(b) + H_{int}(c_1, c_2, c_3). 
\end{equation}
In the next Section  we get the explicit parametrizations of sets quantum states generated by evolution of 
selected initial states and obtain relevant manifolds of dimension depending on the initial states. Then we shall describe 
Riemannian geometry of obtained manifolds introducing the Fubini-Study metrics for each case. Then we study changes of 
geometry resulting from small perturbation in the original Hamiltonian by switching on an additional weak magnetic field 
along the first-axis. Furthermore, in the Section 5, we discuss the characterization of entanglement for each of the 
obtained manifolds using the squared concurrence.      
%
\section{The unitary transformation of two-qubit state \label{sec2}}

Let  a unitary evolution be defined by the Hamiltonian  $H$ given by Eq. (\ref{ham}) describing a two-qubit system with the anisoropic Heisenberg type Hamiltonian in the magnetic field directed along the $z$-axis i.e.
\begin{eqnarray}\label{form1}
H=b\, ( \sigma_3 \otimes \mathds{1}+ \mathds{1}\otimes \sigma_3) + \sum_{j=1}^3 c_j \, \sigma_j \otimes \sigma_j,
\end{eqnarray}
where parameters $c_j$, $j=1,2,3$ are dimensionless interaction couplings between qubits and $b$ is dimensionless parameter decribing an external magnetic field \cite{amf1}. 

This Hamiltonian has four eigenvalues: $E_1^{(0)}=c_3+\omega$, $E_2^{(0)}=c_3-\omega$, $E_3^{(0)}=-c_3 + c_+$, and $E_4^{(0)}=-c_3-c_+$, where  $\omega=\sqrt{(2b)^2+c_-^2}$, $\tan\phi=2b/c_-$ and $c_{\pm}=c_1\pm c_2$. 
The corresponding eigenvectors have the following form
\begin{eqnarray}
&&\vert\psi_1^{(0)}\rangle=\frac{1}{\sqrt{2}}\left[\frac{\cos\phi}{\sqrt{1-\sin\phi}}\vert\uparrow\uparrow\rangle+\sqrt{1-\sin\phi}\vert\downarrow\downarrow\rangle\right],\label{form2_1}\\
&&\vert\psi_2^{(0)}\rangle=\frac{1}{\sqrt{2}}\left[\frac{\cos\phi}{\sqrt{1+\sin\phi}}\vert\uparrow\uparrow\rangle-\sqrt{1+\sin\phi}\vert\downarrow\downarrow\rangle\right],\label{form2_2}\\
&&\vert\psi_3^{(0)}\rangle=\frac{1}{\sqrt{2}}\left(\vert\uparrow\downarrow\rangle + \vert\downarrow\uparrow\rangle\right),\label{form2_3}\\
&&\vert\psi_4^{(0)}\rangle=\frac{1}{\sqrt{2}}\left(\vert\uparrow\downarrow\rangle - \vert\downarrow\uparrow\rangle\right).
\label{form2_4}
\end{eqnarray}
The unitary transformation $U(b, c_j)=U(\omega, \phi, c_{\pm})$ generated by the Hamiltonian (\ref{form1}) acts on
an arbitrary quantum state of two qubits  
\begin{equation}
\vert\psi_I^{(0)}\rangle=\eta_1\vert\psi_1^{(0)}\rangle\nonumber+\eta_2\vert\psi_2^{(0)}\rangle\nonumber+\eta_3\vert\psi_3^{
(0)}\rangle\nonumber+\eta_4\vert\psi_4^{(0)}\rangle\nonumber
\end{equation}
as follows
\begin{equation}
\vert\psi^{(0)}(\omega, \phi, c_{\pm})\rangle=U(\omega, \phi, c_{\pm})\vert\psi_I^{(0)}\rangle
\end{equation}
\begin{eqnarray}\label{form4}
\vert\psi^{(0)}\rangle=&&e^{-ic_3}(\eta_1e^{-i\omega}\vert\psi_1^{(0)}\rangle+\eta_2e^{i\omega}\vert\psi_2^{(0)}\rangle\nonumber\\
&&+\eta_3e^{i(2c_3-c_+)}\vert\psi_3^{(0)}\rangle+\eta_4e^{i(2c_3+c_+)}\vert\psi_4^{(0)}\rangle),
\end{eqnarray}
where the parameters which define the initial state satisfy the normalization condition
$\vert\eta_1\vert^2+\vert\eta_2\vert^2+\vert\eta_3\vert^2+\vert\eta_4\vert^2=1$.

The state (\ref{form4}) depends on four parameters (($\omega$, $\phi$, $c_3$, $c_+$)  satisfying some periodic 
conditions. These conditions, in turn, depend on the initial coordinates $\eta_j$. 

Let us classify the possible parametrizations as follows:
\begin{enumerate}
\item[C1.] For $\eta_1=\eta_2=0$ and  $\eta_3\neq 0$, $\eta_4\neq 0$, the state (\ref{form4}) takes the form
\begin{eqnarray}
\vert\psi^{(0)}\rangle=e^{ic_3}\left(\eta_3 e^{-ic_+}\vert\psi_3^{(0)}\rangle+\eta_4e^{ic_+}\vert\psi_4^{(0)}\rangle\right).
\label{form4_1}
\end{eqnarray}
It is easy to see that this state depends only on parameters $c_+$ and satisfies the following periodic condition
\begin{eqnarray}
\vert\psi^{(0)}\left(c_++\pi\right)\rangle=-\vert\psi^{(0)}\left(c_+\right)\rangle.
\label{form4_2}
\end{eqnarray}
\item[C2.] For $\eta_3=\eta_4=0$ and $\eta_1=0$ or $\eta_2=0$  we obtain the states (\ref{form2_1}) or (\ref{form2_2}) which depend only on parameters $\phi$ with periodic condition
\begin{eqnarray}
\vert\psi^{(0)}\left(\phi+2\pi\right)\rangle=\vert\psi^{(0)}\left(\phi\right)\rangle.
\label{form4_4}
\end{eqnarray}
\item[C3.] For $\eta_3=\eta_4=0$ and non-zero $\eta_1$, $\eta_2$,  the family of states is defined by the parameters $\omega$ and $\phi$ as follows
\begin{eqnarray}
\vert\psi^{(0)}\rangle=e^{-ic_3}\left(\eta_1 e^{-i\omega}\vert\psi_1^{(0)}\rangle+\eta_2e^{i\omega}\vert\psi_2^{(0)}\rangle\right).
\label{form4_6}
\end{eqnarray}
with the following periodic conditions
\begin{eqnarray}
&&\vert\psi^{(0)}\left(\omega+\pi,\phi\right)\rangle=-\vert\psi^{(0)}\left(\omega,\phi\right)\rangle,\nonumber\\
&&\vert\psi^{(0)}\left(\omega,\phi+2\pi\right)\rangle=\vert\psi^{(0)}\left(\omega,\phi\right)\rangle.
\label{form4_7}
\end{eqnarray}
\item[C4.] For $\eta_1=0$ or $\eta_2=0$ and $\eta_3=0$ or $\eta_4=0$  the family of states is defined by two parameters
\begin{eqnarray}
\vert\psi^{(0)}\rangle=e^{-i\left(c_3+(-1)^l\omega\right)}\left(\eta_{l} \vert\psi_{l}^{(0)}\rangle+\eta_{j}e^{ic}\vert\psi_{j}^{(0)}\rangle\right),
\label{form4_6a}
\end{eqnarray}
where $c=2c_3+(-1)^j c_++(-1)^{l+1}\omega$. Here $l=1,2$, $j=3,4$. The states satisfy the following periodic conditions
\begin{eqnarray}
&&\vert\psi^{(0)}\left(\phi+2\pi,c\right)\rangle=\vert\psi^{(0)}\left(\phi,c\right)\rangle,\nonumber\\
&&\vert\psi^{(0)}\left(\phi,c+2\pi\right)\rangle=\vert\psi^{(0)}\left(\phi,c\right)\rangle.
\label{form4_7a}
\end{eqnarray}
\item[C5.] If $\eta_1$, $\eta_2$ are non-zero, and $\eta_3=0$ or $\eta_4=0$ then the family of states is defined by three parameters
\begin{eqnarray}
\vert\psi^{(0)}\rangle=e^{-ic_3}\left(\eta_1 e^{-i\omega}\vert\psi_{1}^{(0)}\rangle+\eta_2 e^{i\omega}\vert\psi_{2}^{(0)}\rangle+\eta_{j}e^{ic}\vert\psi_{j}^{(0)}\rangle\right).
\label{form4_8}
\end{eqnarray}
Here $c=2c_3+(-1)^j c_+$. In this case the states satisfy the following periodic conditions
\begin{eqnarray}
&&\vert\psi^{(0)}\left(\omega+\pi,\phi,c+\pi\right)\rangle=-\vert\psi^{(0)}\left(\omega,\phi,c\right)\rangle,\nonumber\\
&&\vert\psi^{(0)}\left(\omega,\phi+2\pi,c\right)\rangle=\vert\psi^{(0)}\left(\omega,\phi,c\right)\rangle,\nonumber\\
&&\vert\psi^{(0)}\left(\omega,\phi,c+2\pi\right)\rangle=\vert\psi^{(0)}\left(\omega,\phi,c\right)\rangle.
\label{form4_9}
\end{eqnarray}
\item[C6.] For $\eta_1=0$ or $\eta_2=0$, and nonvanishing $\eta_3$, $\eta_4$  the family of states is defined by three parameters
\begin{eqnarray}
\vert\psi^{(0)}\rangle=e^{-i\left(c_3+(-1)^{l+1}\omega\right)}\left(\eta_{l}\vert\psi_{l}^{(0)}\rangle+\eta_3 e^{i\left(c-c_+\right)}\vert\psi_{3}^{(0)}\rangle+\eta_4 e^{i\left(c+c_+\right)}\vert\psi_{4}^{(0)}\rangle\rangle\right).
\label{form4_10}
\end{eqnarray}
Here $c=2c_3+(-1)^{l+1} \omega$. In this case we have the following periodic conditions
\begin{eqnarray}
&&\vert\psi^{(0)}\left(\phi+2\pi,c,c_+\right)\rangle=\vert\psi^{(0)}\left(\phi,c,c_+\right)\rangle,\nonumber\\
&&\vert\psi^{(0)}\left(\phi,c+2\pi,c_+\right)\rangle=\vert\psi^{(0)}\left(\phi,c,c_+\right)\rangle,\nonumber\\
&&\vert\psi^{(0)}\left(\phi,c+\pi,c_++\pi\right)\rangle=\vert\psi^{(0)}\left(\phi,c,c_+\right)\rangle.
\label{form4_11}
\end{eqnarray}
\item[C7.] In the general case, when all parameters $\eta_1$, $\eta_2$, $\eta_3$ and $\eta_4$ are non-zero, we have
the state defined by expression (\ref{form4}) with the following periodic conditions
\begin{eqnarray}
&&\vert\psi^{(0)}\left(\omega+\pi,\phi,c_3+\pi/2,c_+\right)\rangle=i\vert\psi^{(0)}\left(\omega,\phi,c_3,c_+\right)\rangle,\nonumber\\
&&\vert\psi^{(0)}\left(\omega+\pi,\phi,c_3,c_++\pi\right)\rangle=-\vert\psi^{(0)}\left(\omega,\phi,c_3,c_+\right)\rangle,\nonumber\\
&&\vert\psi^{(0)}\left(\omega,\phi+2\pi,c_3,c_+\right)\rangle=\vert\psi^{(0)}\left(\omega,\phi,c_3,c_+\right)\rangle,\nonumber\\
&&\vert\psi^{(0)}\left(\omega,\phi,c_3+\pi,c_+\right)\rangle=-\vert\psi^{(0)}\left(\omega,\phi,c_3,c_+\right)\rangle,\nonumber\\
&&\vert\psi^{(0)}\left(\omega,\phi,c_3+\pi/2,c_++\pi\right)\rangle=-i\vert\psi^{(0)}\left(\omega,\phi,c_3,c_+\right)\rangle.
\label{form4_12}
\end{eqnarray}
\end{enumerate}

Analyzing above cases we can conclude that all obtained quantum state manifolds are closed. However, in the first two 
cases quantum state manifold is one-parametric, in the third and fourth cases it is two-parametric, in the fifth and sixth 
case the manifold is defined by three parameters, and in the last case we have the four-parameter manifold. Let us study 
the Fubiny-Study metric of these  manifolds, $\mathcal{M}_{\psi^{(0)}}$
\begin{table}[ht] 
\centering 
  \begin{tabular}{ | c| c | c| }
    \hline
 Case & $dim {\mathcal{M}}_{\vert\psi^{(0)}\rangle}$ & Parameters  \\ \hline
    C1 & 1 & $c_+$  \\ \hline
    C2 & 1 & $\phi$ \\  \hline
    C3 & 2 &  $\omega$, $\phi$ \\  \hline
    C4 & 2 &  $\phi$, $c$\\  \hline
    C5 & 3 &  $\omega$, $\phi$, $c$ \\  \hline
    C6 & 3 & $\omega$, $c_+$, $c$ \\ \hline
    C7 & 4 &  $\omega$, $\phi$, $c_3$ $c_+$\\ \hline
  \end{tabular}
  \caption{Dimensions and parametrization of the state manifolds.}  
  \label{table:dim} 
  \end{table} 

\section{The Fubini-Study metric of quantum state manifolds \label{sec3}}

The Fubini-Study metric is defined by the infinitesimal distance $ds$ between two neighbouring pure quantum states
$\vert\psi (\xi^{\mu})\rangle$ and $\vert\psi (\xi^{\mu}+d\xi^{\mu})\rangle$ \cite{FSM2}
\begin{eqnarray}
ds^2=g_{\mu\nu}d\xi^{\mu}d\xi^{\nu},
\label{form5}
\end{eqnarray}
where $\xi^{\mu}$ is a set of real parameters which define the state $\vert\psi(\xi^{\mu})\rangle$. The components of the metric tensor
$g_{\mu\nu}$ have the form
\begin{eqnarray}
g_{\mu\nu}=\gamma^2\Re\left(\langle\psi_{\mu}\vert\psi_{\nu}\rangle-\langle\psi_{\mu}\vert\psi\rangle\langle\psi\vert\psi_{\nu}\rangle\right),
\label{form6}
\end{eqnarray}
where $\gamma$ is an arbitrary factor which is often chosen to have value of $1$, $\sqrt{2}$ or $2$ and
\begin{eqnarray}
\vert\psi_{\mu}\rangle=\frac{\partial}{\partial\xi^{\mu}}\vert\psi\rangle.
\label{form7}
\end{eqnarray}

As we have previously noted, the states (\ref{form4}) are defined by four real parameters.
Using definition (\ref{form6}) we obtain the components of the metric tensor with respect to parameters $(\omega,\phi,c_3,c_+)$
\begin{eqnarray}
&&g_{\omega\omega}^{(0)}=\gamma^2\left(\eta_{12}^+-\left(\eta_{12}^-\right)^2\right),\quad g_{\omega\phi}^{(0)}=\gamma^2\eta_{12}^-J,\quad g_{\omega c_3}^{(0)}=2\gamma^2\eta_{12}^-\eta_{34}^+,\nonumber\\
&&g_{\omega c_+}^{(0)}=-\gamma^2\eta_{12}^-\eta_{34}^-,\quad g_{\phi\phi}^{(0)}=\gamma^2\left(\frac{1}{4}\eta_{12}^+-J^2\right),\quad g_{\phi c_3}^{(0)}=-2\gamma^2J\eta_{34}^+\nonumber\\
&&g_{\phi c_+}^{(0)}=\gamma^2J\eta_{34}^-,\quad g_{c_3c_3}^{(0)}=4\gamma^2\eta_{12}^+\eta_{34}^+,\nonumber\\
&&g_{c_3c_+}^{(0)}=-2\gamma^2\eta_{12}^+\eta_{34}^-,\quad g_{c_+c_+}^{(0)}=\gamma^2\left(\eta_{34}^+-\left(\eta_{34}^-\right)^2\right),
\label{form8}
\end{eqnarray}
where $\eta_{ij}^{\pm}=\vert \eta_i\vert^2\pm\vert \eta_j\vert^2$, $J=\Im\left(\eta_1\eta_2^*e^{-2i\omega}\right)$.
From the explicit form of the metric tensor we see, that in the case of the magnetic field switched off,  one of the 
parameters disappears ($\phi=0$) and the the manifold becomes  flat. It is the result of the reciprocal commutativity of
the interaction terms in the Hamiltonian (\ref{form1}).
It is worth noting, that if $c_1=c_2$ and $c_3=\alpha c_+/2$ than $\phi=\pi/2$ and we obtain the metric of the 
two-parameter manifold as in \cite{FMM}
\begin{eqnarray}
&&g_{\omega\omega}^{(0)}=\gamma^2\left(\eta_{12}^+-\left(\eta_{12}^-\right)^2\right),\quad g_{\omega c_+}^{(0)}=2\gamma^2\eta_{12}^-\left(\alpha\eta_{34}^+-\eta_{34}^-\right),\nonumber\\
&&g_{c_+c_+}^{(0)}=\gamma^2\alpha\eta_{12}^+\left(\alpha\eta_{34}^+-2\eta_{34}^-\right)+\gamma^2\left(\eta_{34}^+-\left(\eta_{34}^-\right)^2\right),
\label{form9}
\end{eqnarray}
where $\alpha$ is some real number that determines the anisotropy of the system. If $\alpha=1$ then we obtain the 
Fubini-Study metric of the quantum state manifold of isotropic Heisenberg model \cite{torus}.
The metric (\ref{form8}) can be reduсed to the diagonal form with the use of the new parameters after the following 
transformation
\begin{eqnarray}
\omega=\omega',\quad \phi=k_1\omega'+\phi',\quad c_3=k_2\omega'+k_3\phi'+c_3',\quad c_+=k_4c_3'+c_+',
\label{form9_1}
\end{eqnarray}
where $k_1=4\eta_{12}^-J/\left(4J^2-(\eta_{12}^+)^2\right)$, $k_2=\eta_{12}^+\eta_{12}^-/\left(8J^2-2(\eta_{12}^+)^2\right)$,
$k_3=J/\left(2\eta_{12}^+\right))$, $k_4=2\eta_{12}^+\eta_{34}^-/\left(\eta_{34}^+-(\eta_{34}^-)^2\right)$. Let us 
additionally assume that 
\begin{eqnarray}
J=\frac{\eta_{12}^+}{2}\cos\theta.
\label{form9_2}
\end{eqnarray}
Then in these new parameters the metric (\ref{form8}) takes the following form
\begin{eqnarray}
&&g_{\theta\theta}^{(0)}=\frac{\gamma^2}{4}\eta_{12}^+,\quad g_{\phi'\phi'}^{(0)}=\frac{\gamma^2}{4}\eta_{12}^+\sin^2\theta,\nonumber\\
&&g_{c_3'c_3'}^{(0)}=4\gamma^2\eta_{12}^+\frac{\left(\eta_{34}^+\right)^2-\left(\eta_{34}^-\right)^2}{\eta_{34}^+-\left(\eta_{34}^-\right)^2},\quad g_{c_+'c_+'}^{(0)}=\gamma^2\left(\eta_{34}^+-\left(\eta_{34}^-\right)^2\right).
\label{form9_3}
\end{eqnarray}
It is evident that the ratio between the parameters of the initial state has the influence on the components of the metric 
tensor. For instance if $\eta_{3}=\eta_{4}$ then $\eta_{34}^-=0$ and $g_{c_+'c_+'}$ takes the maximal value for the specific initial state.
Let us analyze in detail the geometry of the manifold defined by above metric for the cases considered in the previous Section:
\begin{enumerate}
\item In the first case the manifold is defined by the parameter $c_+\in\left[0,\pi\right]$ and metric tensor is reduced to $g_{c_+c_+}$
component with $\eta_{34}^+=1$. This is the metric of the circle of the radius $\gamma\sqrt{1-\left(\eta_{34}^{-}\right)^2}/2$.
\item In the second case the manifold is defined by parameter $\phi\in\left[0,2\pi\right]$ and
metric tensor is reduced to $g_{\phi\phi}^{(0)}$ with $\eta_{12}^+=1$. This metric also describes the circle of the radius $\gamma/2$.
\item In this case the manifold is two-parametric $\theta\in\left[0,\pi\right]$, $\phi'\in\left[0,2\pi\right]$
and is described by the metric tensor with components $g_{\theta\theta}^{(0)}$, $g_{\phi'\phi'}^{(0)}$, where $\eta_{12}^+=1$.
This means that it is the sphere of radius $\gamma/2$.
\item Here we have also two-parametric manifold defined by parameters $\phi\in\left[0,2\pi\right]$, $c\in\left[0,2\pi\right]$ and
described by the following metric tensor
\begin{eqnarray}
g_{\phi \phi}^{(0)}=\frac{\gamma^2}{4}\vert\eta_l\vert^2,\quad g_{\phi c}^{(0)}=0,\quad g_{c c}^{(0)}=9\gamma^2\vert\eta_l\vert^2\vert\eta_j\vert^2.
\label{form8_1}
\end{eqnarray}
As we can see that components of the metric tensor do not depend on the parameters $\phi$ and $c$. This means that manifold is flat.
Taking into account periodic conditions (\ref{form4_7}) we conclude that it is a torus.
\item In the fifth case the manifold is three-parametric and defined by the parameters $\theta\in\left[0,\pi\right]$,
$\phi'\in\left[0,2\pi\right]$, $c'\in\left[0,2\pi\right]$. In the diagonal form the metric tensor components $g_{\theta\theta}^{(0)}$, and
$g_{\phi'\phi'}^{(0)}$ are defined by expression (\ref{form9_3}) and other component takes the form
\begin{eqnarray}
g_{c' c'}^{(0)}=4\gamma^2\eta_{12}^+\vert\eta_j\vert^2,
\label{form8_3}
\end{eqnarray}
where $c'$ is related to the parameter $c$ from (\ref{form4_8}) by the following formula
\begin{eqnarray}
c=-\frac{\eta_{12}^-}{2\eta_{12}^+\sin^2\theta}\omega'+\frac{1}{4}\cos\theta\phi'+c'.
\label{form8_4}
\end{eqnarray}
The manifold which we obtain here is the product of the sphere of radius
$\gamma\sqrt{\eta_{12}^+}/2$ in parameters $\theta$, $\phi'$ and of the
circle of radius $2\gamma\sqrt{\eta_{12}^+}\vert\eta_j\vert$ in parameter $c'$.
\item In the case C6 we obtain a manifold with the metric tensor in the diagonal form  
\begin{eqnarray}
g_{\phi \phi}^{(0)}=\frac{\gamma^2}{4}\vert\eta_l\vert^2,\quad g_{c' c'}^{(0)}=4\gamma^2\vert\eta_l\vert^2\frac{\left(\eta_{34}^+\right)^2-\left(\eta_{34}^-\right)^2}{\eta_{34}^+-\left(\eta_{34}^-\right)^2},
\label{form8_5}
\end{eqnarray}
and the component $g_{c_+'c_+'}^{(0)}$ is defined by the expression (\ref{form9_3}). Therefore we obtain a three-parameter 
manifold defined by
$\phi\in\left[0,2\pi\right]$, $c'\in\left[0,2\pi\right]$, $c_+'\in\left[0\pi\right]$. To diagonalize this metric
we use the following transformation
\begin{eqnarray}
c=c',\quad c_+=\frac{2\vert\eta_l\vert^2\eta_{34}^-}{\eta_{34}^+-\left(\eta_{34}^-\right)^2}c'+c_+',
\label{form8_6}
\end{eqnarray}
where $c$ is defined as for the state (\ref{form4_10}). So, this manifold can be expressed by circle of radius $\gamma\vert\eta_l\vert/2$ in parameter
$\phi$ and torus in parameter $c'$, $c_+'$.
\item In the general case the metric is defined by expression (\ref{form8}) or (\ref{form9_3}). This manifold consists of two submanifolds,
namely, the sphere of radius $\gamma\sqrt{\eta_{12}^+}/2$ in parameters $\theta\in\left[0,\pi\right]$, $\phi'\in\left[0,2\pi\right]$ and
torus in parameter $c_3'\in\left[0,\pi\right]$, $c_+'\in\left[0,\pi\right]$.
\end{enumerate}

\section{The Fubini-Study metric of quantum state manifold with perturbation \label{sec4}}

In this Section we study the Fubini-Study metric of quantum state manifold obtained as result of actions of unitary 
transformations generated by the Hamiltonian (\ref{form1}) modified by additional perturbation term which do not 
commute with the original Hamiltonian. This perturbation switches on a weak megnetic field directed along the $x$-axis 
with value $\beta$, where $\beta$ is assumed to be small. Explicitly, the Hamiltonian  of this system takes the form
\begin{eqnarray}
H'=H+\beta\left(\sigma_x\otimes I+I\otimes\sigma_x\right).
\label{form10}
\end{eqnarray}
The eigenvalues and eigenstates of this Hamiltonian we find using the perturbation teory with respect to the first order of $\beta$.
So, the eigenvalues of Hamiltonian (\ref{form10}) are the same as for Hamiltonian (\ref{form1}) which correspond to the following eigenstates
\begin{eqnarray}
&&\vert\psi_1\rangle=\vert\psi_1^{(0)}\rangle+\frac{1}{\sqrt{2}}\frac{\beta}{2c_3+\omega-c_+}\frac{\cos\phi-\sin\phi+1}{\sqrt{1-\sin\phi}}\left[\vert\uparrow\downarrow\rangle+\vert\downarrow\uparrow\rangle\right],\label{form10_1}\\
&&\vert\psi_2\rangle=\vert\psi_2^{(0)}\rangle+\frac{1}{\sqrt{2}}\frac{\beta}{2c_3-\omega-c_+}\frac{\cos\phi-\sin\phi-1}{\sqrt{1+\sin\phi}}\left[\vert\uparrow\downarrow\rangle+\vert\downarrow\uparrow\rangle\right],\label{form10_2}\\
&&\vert\psi_3\rangle=\vert\psi_3^{(0)}\rangle+\frac{\beta}{-2c_3-\omega+c_+}\frac{\cos\phi-\sin\phi+1}{\sqrt{2}}
\left[\frac{\cos\phi}{1-\sin\phi}\vert\uparrow\uparrow\rangle+\vert\downarrow\downarrow\rangle\right]\nonumber\\
&&+\frac{\cos\phi-\sin\phi-1}{\sqrt{2}}\frac{\beta}{-2c_3+\omega+c_+}\left[\frac{\cos\phi}{1+\sin\phi}\vert\uparrow\uparrow\rangle-\vert\downarrow\downarrow\rangle\right],\label{form10_3}\\
&&\vert\psi_4\rangle=\vert\psi_4^{(0)}\rangle.
\label{form10_4}
\end{eqnarray}
Similarly to the previous case we decompose the initial state using the eigenstates (\ref{form10_4}). Then the evolution
is described by equation (\ref{form4}) with above eigenstates.
the Fubini-Study metric for evolution generated by the linearly perturbed Hamiltonian takes the following form
\begin{eqnarray}
&&g_{\omega \omega} = g_{\omega \omega}^{(0)}
+ 2 \beta \gamma^2 \left\lbrace(1 - 2 \eta_{12}^-) \Im\left(\eta_1 \eta_3^* e^{- i (2 c_3 + \omega - c_+)}\right) Y_+\right.\nonumber\\
&&\left.+ (1 + 2 \eta_{12}^-) \Im\left(\eta_2 \eta_3^* e^{-i (2 c_3   - \omega- c_+)}\right) Y_-\right\rbrace,
\label{form10_5}
\end{eqnarray}
\begin{eqnarray}
&&g_{c_3 c_3} = g_{c_3 c_3}^{(0)}  - 8 \beta \gamma^2
(\eta_{12}^+ - \eta_{34}^+)
\left\lbrace
\Im\left(\eta_1 \eta_3^* e^{-i (2 c_3 + \omega - c_+)}\right) Y_+\right.\nonumber\\
&&\left.+ \Im\left(\eta_2 \eta_3^* e^{-i (2 c_3 - \omega - c_+)}\right) Y_-
\right\rbrace,
\end{eqnarray}
\begin{eqnarray}
&&g_{c_+ c_+} = g_{c_+ c_+}^{(0)}   - 2 \beta \gamma^2
(1 - 2 \eta_{34}^-)
\left\lbrace
\Im\left(\eta_1 \eta_3^* e^{-i (2 c_3 + \omega - c_+)}\right) Y_+\right.\nonumber\\
&&\left.+ \Im\left(\eta_2 \eta_3^* e^{ -i (2 c_3 - \omega - c_+)}\right) Y_-
\right\rbrace,
\end{eqnarray}
\begin{eqnarray}
&&g_{\phi \phi} = g_{\phi \phi}^{(0)}
+ \beta \gamma^2\omega \left\lbrace\left(4J\Im\left(\eta_1\eta_3^*e^{-i(2c_3+\omega-c_+)}\right)-\Re\left(\eta_2\eta_3^*e^{-i(2c_3-\omega-c_+)}\right)\right)X_-\right.\nonumber\\
&&\left.+\left(4J\Im\left(\eta_2\eta_3^*e^{-i(2c_3-\omega-c_+)}\right)+\Re\left(\eta_1\eta_3^*e^{-i(2c_3+\omega-c_+)}\right)\right)X_+
\right\rbrace,
\end{eqnarray}
\begin{eqnarray}
&&g_{\phi \omega} = g_{\phi \omega}^{(0)}
+ \beta \gamma^2\left\lbrace
\omega\left(1-\eta_{12}^-\right)\Im\left( \eta_1 \eta_3^* e^{-i (2 c_3 + \omega - c_+)} \right) X_-\right.\nonumber\\
&&-\omega\left(1+\eta_{12}^-\right)\Im\left( \eta_2 \eta_3^* e^{-i (2 c_3 - \omega - c_+)} \right) X_+\nonumber\\
&&-\left(\frac{1}{2}\Re \left( \eta_1 \eta_3^* e^{-i (2 c_3 + \omega - c_+)}\right) + 2 J \Im\left( \eta_2 \eta_3^* e^{-i (2 c_3 - \omega - c_+)} \right)\right)Y_-\nonumber\\
&&\left.- \left(\frac{1}{2}\Re \left( \eta_2 \eta_3^* e^{-i (2 c_3 - \omega - c_+)}\right) - 2 J \Im\left( \eta_1 \eta_3^* e^{-i (2 c_3 + \omega - c_+)} \right)\right)Y_+
\right\rbrace,
\end{eqnarray}
\begin{eqnarray}
&&g_{c_3 \omega} = g_{c_3 \omega}^{(0)}
+ 4 \beta \gamma^2\left\lbrace
\left(\eta_{34}^+-\eta_{12}^-\right)\Im \left( \eta_1 \eta_3^* e^{-i (2 c_3 + \omega - c_+)} \right)Y_+\right.\nonumber\\
&&\left.-\left(\eta_{34}^++\eta_{12}^-\right)\Im \left( \eta_2 \eta_3^* e^{-i (2 c_3 - \omega - c_+)} \right)Y_-
\right\rbrace,
\end{eqnarray}
\begin{eqnarray}
&&g_{c_+ \omega} = g_{c_+ \omega}^{(0)}
+ 2 \beta \gamma^2\left\lbrace
\left(\eta_{12}^--\eta_{34}^-\right)\Im \left( \eta_1 \eta_3^* e^{-i (2 c_3 + \omega - c_+)} \right)Y_+\right.\nonumber\\
&&\left.+\left(\eta_{12}^-+\eta_{34}^-\right)\Im \left( \eta_2 \eta_3^* e^{-i (2 c_3 - \omega - c_+)} \right)Y_-
\right\rbrace,
\end{eqnarray}
\begin{eqnarray}
&&g_{c_3 \phi} = g_{c_3 \phi}^{(0)}\nonumber\\
&&+\beta \gamma^2\left\lbrace -2\omega\left(\eta_{12}^+-\eta_{34}^+\right)\left(\Im\left(\eta_1\eta_3^*e^{-i(2c_3+\omega-c_+)}\right)X_-
+\Im\left(\eta_2\eta_3^*e^{-i(2c_3-\omega-c_+)}\right)X_+\right)\right.\nonumber\\
&&+\left(\Re\left(\eta_1\eta_3^*e^{-i(2c_3+\omega-c_+)}\right)+4J\Im\left(\eta_2\eta_3^*e^{-i(2c_3-\omega-c_+)}\right)\right)Y_-\nonumber\\
&&\left.-\left(\Re\left(\eta_2\eta_3^*e^{-i(2c_3-\omega-c_+)}\right)-4J\Im\left(\eta_1\eta_3^*e^{-i(2c_3+\omega-c_+)}\right)\right)Y_+\right\rbrace
\end{eqnarray}
\begin{eqnarray}
&&g_{c_+ \phi} = g_{c_+ \phi}^{(0)}\nonumber\\
&&+\beta \gamma^2\left\lbrace \omega\left(1-2\eta_{34}^-\right)\left(\Im\left(\eta_1\eta_3^*e^{-i(2c_3+\omega-c_+)}\right)X_-
+\Im\left(\eta_2\eta_3^*e^{-i(2c_3-\omega-c_+)}\right)X_+\right)\right.\nonumber\\
&&-\left(\frac{1}{2}\Re\left(\eta_1\eta_3^*e^{-i(2c_3+\omega-c_+)}\right)+2J\Im\left(\eta_2\eta_3^*e^{-i(2c_3-\omega-c_+)}\right)\right)Y_-\nonumber\\
&&\left.+\left(\frac{1}{2}\Re\left(\eta_2\eta_3^*e^{-i(2c_3-\omega-c_+)}\right)-2J\Im\left(\eta_1\eta_3^*e^{-i(2c_3+\omega-c_+)}\right)\right)Y_+\right\rbrace
\end{eqnarray}
\begin{eqnarray}
&&g_{c_+ c_3} = g_{c_+ c_3}^{(0)}   + 2 \beta \gamma^2
(2\eta_{12}^+ - \eta_{34}^-)
\left\lbrace
\Im\left(\eta_1 \eta_3^* e^{-i (2 c_3 + \omega - c_+)}\right) Y_+\right.\nonumber\\
&&\left.+ \Im\left(\eta_2 \eta_3^* e^{ -i (2 c_3 - \omega - c_+)}\right) Y_-
\right\rbrace,
\end{eqnarray}
where we use the following notation
\begin{eqnarray}
Y_{\pm}=\frac{\sqrt{1 - \sin(\phi)}\pm\sqrt{1 + \sin(\phi)}}{(2 c_3 - c_+ \pm \omega)^2},\quad
X_{\pm}=\frac{\sqrt{1-\sin(\phi)}\pm\sqrt{1+\sin(\phi)}}{(2c_3-c_+)^2-\omega^2}.\nonumber
\end{eqnarray}
and above components give the new perturbed metric of the form
\begin{equation}\label{pertmet}
g_{ij}=g^{(0)} _{ij}+ \beta h_{ij}
\end{equation}
Such perturbation modifies geometry of some state manifolds enlisted in the Table 1. For the cases C1 to C3 there is no 
modification at all. The case C4 for choices $\eta_2 \neq 0$ or $\eta_4 \neq 0$ is also unperturbed. A nontrivial modification
appears for C5, C6 and C7. The explicit formulas for the scalar curvature of perturbed metric for these cases are hard to obtain. 
To illustrate the effect of modification of the manifold let us consider special initial conditions for the case C7. Let us assume 
that $\eta_1=\eta_2=\eta_3=\eta_4=\frac{1}{2}$. Then the metric (\ref{form8})   takes the following form
\begin{equation}
G^0 = \left(
\begin{array}{cccc}
	\frac{\gamma ^2}{2} & 0 & 0 & 0 \\
	0 & \frac{1}{32} \gamma ^2 (\cos (2 (\alpha_{12}+2 \omega ))+3) & \frac{1}{4} \gamma ^2 \sin (\alpha_{12}+2 \omega ) & 0 \\
	0 & \frac{1}{4} \gamma ^2 \sin (\alpha_{12}+2 \omega ) & \gamma ^2 & 0 \\
	0 & 0 & 0 & \frac{\gamma ^2}{2} \\
\end{array}
\right)
\end{equation} 
and yields the Ricci tensor
\begin{equation}
\mathcal{R} = 
\left(
\begin{array}{cccc}
	3 & 0 & 0 & 0 \\
	0 & \frac{1}{16} (5 \cos (2 (\text{$\alpha_{12}$}+2 \omega ))+7) & \frac{1}{2} \sin (\text{$\alpha_{12}$}+2 \omega ) & 0 \\
	0 & \frac{1}{2} \sin (\text{$\alpha_{12}$}+2 \omega ) & 2 & 0 \\
	0 & 0 & 0 & 0 \\
\end{array}
\right)
\end{equation} 
where $\alpha_{ij}=\alpha_i-\alpha_j$ and the scalar curvature $R=\frac{14}{\gamma^2}$. Here, parameters $\phi$, $c_+$, $c_3$, $\alpha_{13}$ and
$\alpha_{23}$  do not influence the value of $R$, one can also put $\alpha_{12}=0$. Now for the perturbed metric (\ref{pertmet})  depending solely on $\omega$ we get the following scalar curvature
\begin{equation}\label{pertcur}
 R = \frac{\cos (2 \omega )}{\gamma ^2 (\cos (4 \omega )+1)^2} 
  \left(
  \frac{A_1}{A_2}+\frac{B_1}{B_2}+\frac{C_1}{C_2}+\frac{D_1}{D_2}+\frac{E_1}{E_2}   
  \right) 
\end{equation}
where functions entering above formula have the form 
\begin{equation}
	A_1 = 8 \beta  \cos ^4(2 \omega ) \left(4 \omega ^2 \sin (\omega )+6 \left(\omega ^2-2\right) \sin \omega \cos (2 \omega )+8 \omega  \cos (3 \omega )\right)
\end{equation}
\begin{eqnarray} \nonumber
A_2&=&6 \beta  \omega ^2 (4 \omega  \sin (\omega )+\sin (\omega )+\sin (3 \omega )-4 \omega  \cos (\omega )+\cos (\omega )\\  
&&-\cos (5 \omega ))+\omega ^4+\omega ^4 \cos (4 \omega )
\end{eqnarray}
\begin{eqnarray} \nonumber
B_1 &=&(\cos (4 \omega )+1) (\beta (-16 \omega ^2 \sin (\omega )+16 \omega ^2 \sin (3 \omega )
\\\nonumber
&&
-2 (8 \omega ^2+3 \omega -12) \cos (\omega )-4 (4 \omega ^2+9 \omega -1) \cos (3 \omega )
\\\nonumber
&&
+5 \omega  \sin (\omega )+5 \omega  \sin (3 \omega )+\omega  \sin (5 \omega )
+\omega  \sin (7 \omega ) -4 \sin (\omega)\\\nonumber
&&
+24 \sin (3 \omega )+4 \sin (7 \omega )-6 \omega  \cos (7 \omega )+4 \cos (5 \omega ))
\\ 
&&
+7 \omega ^3 \cos (2 \omega )+\omega ^3 \cos (6 \omega ))
\end{eqnarray}
\begin{eqnarray} \nonumber
B_2&=& 4 \beta  \omega  (4 \omega  \sin (\omega )+\sin (\omega )+\sin (3 \omega )-4 \omega  \cos (\omega )+\cos (\omega )\\
&&-\cos (5 \omega ))+\omega ^3+\omega ^3 \cos (4 \omega )
\end{eqnarray}
\begin{eqnarray} \nonumber
\end{eqnarray}
\begin{eqnarray} \nonumber
C_1 &=&  2 \sin (4 \omega ) (2 \beta  \left(14 \omega ^3+4 \omega ^2-7 \omega +6\right) \cos (\omega )-2 \beta  (14 \omega ^3 \sin (\omega )\\ 
&&+10 \omega ^3 \sin (3 \omega )-12 \omega ^2 \sin (\omega )+7 \omega ^2 \sin (3 \omega )+7 \omega ^2 \sin (7 \omega )\\\nonumber
&&+\omega ^2 \cos (7 \omega )+(4 \omega ^2-3) \cos (5 \omega )+(10 \omega ^3+5 \omega ^2+12 \omega -3) \cos (3 \omega )\\\nonumber
&&-\omega  \sin (\omega )-\omega  \sin (7 \omega )+3 \sin (\omega )-6 \sin (3 \omega )-3 \sin (7 \omega )\\ \nonumber
&&+5 \omega \cos(7 \omega ))+\omega ^4 \sin (2 \omega )+\omega ^4 \sin(6 \omega )) \nonumber
\end{eqnarray}
\begin{eqnarray} 
C_2&=& 6 \beta  \omega ^2 (4 \omega  \sin (\omega )+\sin (\omega )+\sin (3 \omega )-4 \omega  \cos (\omega )+\cos (\omega )\\\nonumber
&&-\cos (5 \omega ))+\omega ^4+\omega ^4 \cos (4 \omega ) \nonumber
\end{eqnarray}
\begin{eqnarray} \nonumber
D_1&=& \cos (2 \omega ) (\beta  (160 \omega ^2 \sin (\omega )-152 \omega ^2 \sin (3 \omega )+104 \omega ^2 \sin (5 \omega )-104 \omega ^2 \cos (5 \omega )\hfill\\\nonumber
&&-2 \left(80 \omega ^2+9 \omega -32\right) \cos (\omega )+\left(-152 \omega ^2+62 \omega +40\right) \cos (3 \omega )+52 \omega  \sin (\omega )\\\nonumber
&&-46 \omega  \sin (3 \omega )+42 \omega  \sin (5 \omega )+19 \omega  \sin (7 \omega )+7 \omega  \sin (9 \omega )+16 \sin (3 \omega )\\ 
&&+16 \sin (5 \omega )-54 \omega  \cos (5 \omega )-22 \omega  \cos (9 \omega )+24 \cos (5 \omega )-4 \cos (7 \omega )\\\nonumber
&&+4 \cos (9 \omega ))+19 \omega ^3+24 \omega ^3 \cos (4 \omega )
+5 \omega ^3 \cos (8 \omega ))\hfill \nonumber
\end{eqnarray}
\begin{eqnarray} \nonumber
D_2&=&6 \beta  \omega  (4 \omega  \sin (\omega )+\sin (\omega )+\sin (3 \omega )-4 \omega  \cos (\omega )+\cos (\omega )-\cos (5 \omega ))\\ 
&&+\omega ^3+\omega ^3 \cos (4 \omega )\hfill 
\end{eqnarray}
\begin{eqnarray} \nonumber
E_1&=& \cos (2 \omega ) (-2 \beta  \left(40 \omega ^3+21 \omega ^2-44 \omega +6\right) \cos (\omega )+2 \beta  (-100 \omega ^3+55 \omega ^2\\\nonumber
&&+36 \omega -6) \cos (3 \omega )+\beta  (80 \omega ^3 \sin (\omega )-200 \omega ^3 \sin (3 \omega)+136 \omega ^3 \sin (5 \omega) \\\nonumber
&&+24 \omega ^2 \sin (\omega )-82 \omega ^2 \sin (3 \omega )+54 \omega ^2 \sin (5 \omega )\\
&&+25 \omega ^2 \sin (7 \omega )+9 \omega ^2 \sin (9 \omega )+(-38 \omega ^2+8 \omega +12) \cos (9 \omega )\\\nonumber
&&-2 (68 \omega ^3+31 \omega ^2-12 \omega -6) \cos (5 \omega )-16 \omega  \sin (\omega )+48 \omega  \sin (3 \omega )\\\nonumber
&&+48 \omega  \sin (5 \omega )+16 \omega  \sin (9 \omega )-48 \sin (\omega )+12 \sin (3 \omega )-36 \sin (5 \omega )\\
&&-6 \sin (7 \omega )-6 \sin (9 \omega ))+18 \omega ^4+24 \omega ^4 \cos (4 \omega )+6 \omega ^4 \cos (8 \omega ))\hfill \nonumber
\end{eqnarray}
\begin{eqnarray} \nonumber
E_2&=& 6 \beta  \omega ^2 (4 \omega  \sin (\omega )+\sin (\omega )+\sin (3 \omega )-4 \omega  \cos (\omega )+\cos (\omega )-\cos (5 \omega ))\\
&&+\omega ^4+\omega ^4 \cos (4 \omega )\hfill
\end{eqnarray}
Let us note that switching off the perturbation (\ref{pertcur}) gives the correct unperturbed scalar curvature.
%

%
\section{Entanglement characterization of two-qubit quantum state manifolds ${\mathcal{M}}_{\vert\psi^{(0)}\rangle}$ }

In the present section, using the squared concurrence as an entanglement measure,  we shall study the entanglement of states belonging to the manifolds obtained  in the Section \ref{sec2}. The concurrence of a pure state of bipartite two-level system  is defined as follows \cite{ent1,ent2}
\begin{eqnarray}
C(\vert \psi \rangle)=2\vert ad-bc\vert,
\label{form11}
\end{eqnarray}
where $a$, $b$, $c$ and $d$ are defined by expression
\begin{eqnarray}
\vert \psi \rangle = a\vert\uparrow\uparrow\rangle+b\vert\uparrow\downarrow\rangle+c\vert\downarrow\uparrow\rangle+d\vert\downarrow\downarrow\rangle .
\label{form12a}
\end{eqnarray}
The squared concurrence for state (\ref{form4}) takes the form
\begin{eqnarray}
C=\vert \left(\eta_1^2e^{-2i\omega}-\eta_2^2e^{2i\omega}\right)\cos\phi-2\eta_1\eta_2\sin\phi-e^{4ic_3}\left(\eta_3^2e^{-2ic_+}-\eta_4^2e^{2ic_+}\right) \vert .
\label{form12b}
\end{eqnarray}
Let us calculate the squared concurrence for the families of states discussed in the previous sections:
\begin{enumerate}
\item In the case C1 the concurrence takes the form
\begin{eqnarray}
C=\sqrt{\vert\eta_3\vert^4+\vert\eta_4\vert^4-2\vert\eta_3\vert^2\vert\eta_4\vert^2\cos\left(4c_++2\chi\right)}.
\label{form13}
\end{eqnarray}
For $\eta_3=\vert\eta_3\vert$ and $\eta_3=\vert\eta_3\vert e^{i\chi}$, where $\chi\in\left[0,2\pi\right]$. 
we obtain the maximally entangled state if $c_+=1/4\left[(2n+1)\pi-2\chi\right]$, where $n\in\mathds{Z}$.
\item  The squared concurrence in the case C2 takes simple form
\begin{eqnarray}
C=\vert\cos\phi\vert .
\label{form14}
\end{eqnarray}
\item For the C3 family of states the manifold is defined by two parameters. The entanglement of the states is described by the following expression
\begin{eqnarray}
&&C=\left[\left(\vert\eta_1\vert^4+\vert\eta_2\vert^4-2\vert\eta_1\vert^2\vert\eta_2\vert^2\cos\left(4\omega+2\chi\right)\right)\cos^2\phi+4\vert\eta_1\vert^2\vert\eta_2\vert^2\sin^2\phi\right.\nonumber\\
&&\left.-4\vert\eta_1\vert\vert\eta_2\vert\left(\vert\eta_1\vert^2-\vert\eta_2\vert^2\right)\cos\left(2\omega+\chi\right)\sin\phi\cos\phi\right]^{1/2}.
\label{form15}
\end{eqnarray}
Similarly as in the previous case C1 we put $\eta_1=\vert\eta_1\vert$ and $\eta_2=\vert\eta_2\vert e^{i\chi}$.
As we can see, regardless of the initial state the maximally entangled state
we obtain when $\phi=0$ and $\omega=1/4\left[(2n+1)\pi-2\chi\right]$.
\item Here we also put $\eta_l=\vert\eta_l\vert$ and $\eta_j=\vert\eta_j\vert e^{i\chi}$ and obtain the expression for concurrence
\begin{eqnarray}
C=\sqrt{\vert\eta_l\vert^4\cos^2\phi+\vert\eta_j\vert^4-2(-1)^{l+j}\vert\eta_l\vert^2\vert\eta_j\vert^2\cos\left(2c+\chi\right)\cos\phi}.
\label{form16}
\end{eqnarray}
So, the conditions for maximally entangled is the following: $\phi=0$ and $c=1/2\left[(2n+1)\pi-\chi\right]$ for even $l+j$,
$c=1/2\left[2\pi n-\chi\right]$ for odd $l+j$.
\item Fort the C5-family of states, to simplify the calculations, we analyze the case when $\eta_1=\eta_2$ and we put $\eta_1=\vert\eta_1\vert$, $\eta_j=\vert\eta_j\vert e^{i\chi}$. The squared concurrence takes finally the form
\begin{eqnarray}
&&C=\left[\left(-2\vert\eta_1\vert^2\sin\phi+(-1)^j\vert\eta_j\vert^2\cos\left(2c+2\chi\right)\right)^2\right.\nonumber\\
&&\left.+\left(-2\vert\eta_1\vert^2\sin\omega\cos\phi+(-1)^j\vert\eta_j\vert^2\sin\left(2c+2\chi\right)\right)^2\right]^{1/2}.
\label{form17}
\end{eqnarray}
The conditions for peparation of maximally entangled states are given in Table \ref{con1}.
\begin{table}[ht] 
\centering 
  \begin{tabular}{ | c| c | c| c| }
    \hline
 $\phi$ & $\omega$                   & $j$            & $c$                          \\ \hline
\multirow{2}{*}{$0$} & \multirow{2}{*}{$\pi/2$}   & \multirow{1}{*}{even}           & $3\pi/4+\pi n-\chi$          \\
                    &                 & odd            & $\pi/4+\pi n-\chi$           \\ \hline
\multirow{2}{*}{$\pi/2$} & \multirow{2}{*}{--}        & \multirow{1}{*}{even}           & $1/2\left[(2n+1)\pi-\chi\right]$  \\
                    &                 & odd            & $\pi n-\chi$                 \\  \hline
\multirow{2}{*}{$\pi$}   & \multirow{2}{*}{$\pi/2$}   & \multirow{1}{*}{even}           & $\pi/4+\pi n-\chi$           \\
                    &                 & odd            & $3\pi/4+\pi n-\chi$          \\ \hline
\multirow{2}{*}{$3\pi/2$}& \multirow{2}{*}{--}        & \multirow{1}{*}{even}           & $\pi n -\chi$  \\
                    &                 & odd            & $1/2\left[(2n+1)\pi-\chi\right]$         \\  \hline
  \end{tabular}
  \caption{Conditions for maximally entangled states in case C5.}
  \label{con1} 
  \end{table} 
\item We shall use imilar simplifications in the case C6. Here, we also put $\eta_3=\eta_4$ and $\eta_l=\vert\eta_l\vert$, $\eta_3=\vert\eta_3\vert e^{i\chi}$. Then the squared concurrence takes the form
\begin{eqnarray}
&&C=\left[\left((-1)^{l+1}\vert\eta_l\vert^2\cos\phi-2\vert\eta_3\vert^2\sin\left(2c+2\chi\right)\sin 2c_+\right)^2\right.\nonumber\\
&&\left.+4\vert\eta_3\vert^4\cos^2\left(2c+2\chi\right)\sin^22c_+ \right]^{1/2}.
\label{form18}
\end{eqnarray}
The conditions for peparation of maximally entangled states for the C6-family are presented in the Table \ref{con2}.
\begin{table}[ht] 
\centering 
  \begin{tabular}{ | c| c | c| c| }
    \hline
 $\phi$              & $l$                     & $c$                             & $c_+$                          \\ \hline
\multirow{4}{*}{$0$} & \multirow{2}{*}{even}   & $\pi/4+\pi n-\chi$              & $\pi/4+\pi n$          \\
                     &                         & $3\pi/4+\pi n-\chi$             & $3\pi/4+\pi n$          \\
                     & \multirow{2}{*}{odd}    & $\pi/4+\pi n-\chi$              & $3\pi/4+\pi n$          \\
                     &                         & $3\pi/4+\pi n-\chi$             & $\pi/4+\pi n$           \\ \hline
\multirow{4}{*}{$\pi$}& \multirow{2}{*}{even}  & $\pi/4+\pi n-\chi$              & $3\pi/4+\pi n$          \\
                     &                         & $3\pi/4+\pi n-\chi$             & $\pi/4+\pi n$          \\
                     & \multirow{2}{*}{odd}    & $\pi/4+\pi n-\chi$              & $\pi/4+\pi n$          \\
                     &                         & $3\pi/4+\pi n-\chi$             & $3\pi/4+\pi n$           \\ \hline
  \end{tabular}
  \caption{Conditions for  maximally entangled states in the case C6.}
  \label{con2}
  \end{table}
\item In the C7 case we assume that $\eta_1=\eta_2=\vert\eta_2\vert$, $\eta_3=\eta_4=\vert\eta_3\vert e^{i\chi}$ what yelds the squared concurrence in the form
\begin{eqnarray}
&&C=\left[\left(2\vert\eta_1\vert^2\sin\phi+2\vert\eta_3\vert^2\sin 2c_+\sin\left(4c_3+2\chi\right)\right)^2\right.\nonumber\\
&&\left.+\left(-2\vert\eta_1\vert^2\sin 2\omega\cos\phi+2\vert\eta_3\vert^2\sin 2c_+\cos\left(4c_3+2\chi\right)\right)^2\right]^{1/2}
\label{form19}
\end{eqnarray}
The conditions definig maximally entangled states are collected in the Table \ref{con3}
\begin{table}[ht] 
\centering 
  \begin{tabular}{ | c| c | c| c| }
    \hline
 $\phi$              & $\omega$                & $c_+$                           & $c_3$                          \\ \hline
\multirow{4}{*}{$0$} & \multirow{2}{*}{$\pi/4$}& $\pi/4+\pi n$                   & $1/4\left[(2n+1)\pi -2\chi\right]$          \\
                     &                         & $3\pi/4+\pi n$                  & $1/2\left[\pi n-\chi\right]$          \\
                     & \multirow{2}{*}{$3\pi/4$}& $\pi/4+\pi n$                  & $1/2\left[\pi n-\chi\right]$          \\
                     &                         & $3\pi/4+\pi n$                  & $1/4\left[(2n+1)\pi -2\chi\right]$           \\ \hline
\multirow{2}{*}{$\pi/2$}& \multirow{2}{*}{--}  & $\pi/4+\pi n$                   & $1/4\left[\pi/2+2\pi n -2\chi\right]$          \\
                     &                         & $3\pi/4+\pi n$                  & $1/4\left[3\pi/2+2\pi n -2\chi\right]$          \\ \hline
\multirow{4}{*}{$\pi$} & \multirow{2}{*}{$\pi/4$}& $\pi/4+\pi n$                   & $1/2\left[\pi n-\chi\right]$          \\
                     &                         & $3\pi/4+\pi n$                  & $1/4\left[(2n+1)\pi -2\chi\right]$          \\
                     & \multirow{2}{*}{$3\pi/4$}& $\pi/4+\pi n$                  & $1/4\left[(2n+1)\pi -2\chi\right]$          \\
                     &                         & $3\pi/4+\pi n$                  & $1/2\left[\pi n-\chi\right]$           \\ \hline
\multirow{2}{*}{$3\pi/2$}& \multirow{2}{*}{--} & $\pi/4+\pi n$                   & $1/4\left[3\pi/2+2\pi n -2\chi\right]$          \\
                     &                         & $3\pi/4+\pi n$                  & $1/4\left[\pi/2+2\pi n -2\chi\right]$          \\ \hline

  \end{tabular}
  \caption{Conditions for maximally entangled states in case C7.}
  \label{con3}
  \end{table}

\end{enumerate}

\section{Conclusions}
The geometric characterization of the state manifold of quantum system is of great value, but for  compound systems such 
task becomes very complex when addressed in general setting. 

In the present work we have studied quantum state manifolds obtained by 
means of the unitary evolution defined by  large family of physically interesting Hamiltonians.  
Despite the knowledge of the whole set of the two-qubit quantum state space it is important to know what manifolds lying inside this set can be reached using the evolution governed by the realistic Hamiltonians.
The geometry of such obtained quantum state spaces is of Riemannian type defined by the Fubini-Study metrics depending on  initial conditions and parameters entering the definition of the families of Hamiltonians. We have given the classification of possible state manifolds and thoroughly discussed the explicit description of two-qubit unitary orbits generated by physically relevant Hamiltonians. The relevant Fubini-Study metrics were obtained with the use of explicit parametrizations.

It is worth noting, that we also studied the question how obtained geometries 
are modified by the noncommutative linear perturbation term included into the original Hamiltonian. 
We describe its influence on the scalar curvature of the relevant state spaces. In some cases the answer turns out to be nontrivial. 

As an important physical characterization of the considered systems we have studied the degree of entanglement of states for all obtained quantum state spaces  and we have provided conditions for obtainig maximally entangled state in each case, where the concurrence is used as an entanglement monotone.  
\section*{Acknowledgements}
One of the Authors (A.K.) wishes to thank the Institute of Theoretical Physics at the University of Wroclaw for hospitality and financial support as well as  he acknowledges that the work was  supported by Project FF-30F (No. 0116U001539) from the Ministry of Education and Science of Ukraine.
%

\end{document}